\documentclass[fleqn,usenatbib]{mnras}

\usepackage[T1]{fontenc}

\DeclareRobustCommand{\VAN}[3]{#2}
\let\VANthebibliography\thebibliography
\def\thebibliography{\DeclareRobustCommand{\VAN}[3]{##3}\VANthebibliography}

\usepackage{graphicx}	
\usepackage{amsmath}	
\usepackage{amssymb}

\usepackage{graphicx}
\newcommand\SIMI{N2M}
\newcommand\SIMII{N1M}
\newcommand\SIMIII{N05M}
\newcommand\SIMIV{N05M\_wf}
\newcommand\SIMV{N1M\_a}

\usepackage[normalem]{ulem}

\title[Energy Equipartition]{Energy equipartition in multiple-population globular clusters}
\author[A. R. Livernois et al.]{Alexander R. Livernois,$^{1}$\thanks{E-mail: allivern@iu.edu} 
F. I. Aros,$^{1}$ 
E. Vesperini,$^{1}$ 
A. Askar,$^{2}$ 
A. Bellini,$^{3}$ 
\newauthor
M. Giersz,$^{2}$ 
J. Hong,$^{4}$ 
A. Hypki,$^{2,5}$ 
M. Libralato$^{7}$ and
T. Ziliotto$^{8}$
\\
$^{1}$Department of Astronomy, Indiana University, Bloomington, IN, 47405, USA\\
$^{2}$Nicolaus Copernicus Astronomical Center, Polish Academy of Sciences, Bartycka 18, 00-716 Warsaw, Poland\\
$^{3}$Space Telescope Science Institute, 3700 San Martin Drive, Baltimore, MD 21218, USA\\
$^{4}$Korea Astronomy and Space Science Institute, Daejeon 34055, Republic of Korea\\
$^{5}$Faculty of Mathematics and Computer Science, A. Mickiewicz University, Uniwersytetu Poznanskiego 4, 61-614 Poznan, Poland\\
$^{7}$Istituto Nazionale di Astrofisica, Osservatorio Astronomico di Padova, Vicolo dell’Osservatorio 5, Padova I-35122, Italy\\
$^{8}$Dipartimento di Fisica e Astronomia “Galileo Galilei”, Univ. di Padova, Vicolo dell’Osservatorio 3, Padova, IT-35122
}

\graphicspath{{./figures/}}
\begin{document}

\label{firstpage}
\pagerange{\pageref{firstpage}--\pageref{lastpage}}
\maketitle

\begin{abstract}
We present the results of Monte Carlo simulations aimed at exploring the evolution towards energy equipartition of first- (1G) and second-generation (2G) stars in multiple-population globular clusters and how this evolution is affected by the initial differences between the spatial distributions of the two populations.
Our results show that these initial differences  have fundamental implications for the evolution towards energy equipartition of the two populations.
We find that 2G stars, which are assumed to be initially more centrally concentrated than 1G stars, are generally characterized by a more rapid evolution towards energy equipartition. The evolution towards energy equipartition depends on the velocity dispersion component and is more rapid for the tangential velocity dispersion.  The extent of the present-day differences between the degree of energy equipartition of 2G and 1G stars depends on the cluster's dynamical age and may be more significant in the tangential velocity dispersion and at intermediate distances from the cluster's center around the half-mass radius. 
\end{abstract}

\begin{keywords}
globular clusters: general, stars: kinematics and dynamics, methods: numerical
\end{keywords}

\section{Introduction}

Numerous observational studies have shown that globular clusters host  multiple stellar populations characterized by differences in their chemical properties (see e.g. \citealt{2018BaLa}, \citealt{2019GrBr}, \citealt{2022MiMa} for some recent reviews and references therein). 

These studies find that, in addition to a population with chemical properties similar to those of halo field stars with the same metallicity (hereafter referred to as first-generation, 1G), globular clusters host one, or more, groups of chemically anomalous stars (hereafter second-generation, 2G) typically characterized by enhanced Na, Al, N, and helium abundances, and depletion in O, Mg, and C (see e.g. \citealt{2009CaBr2}, \citeyear{2009CaBr}, \citealt{2012gratton} and references therein). 
About 20 per cent of Galactic globular clusters exhibit also significant differences in Fe (see e.g. \citealt{2017MiPi}, \citealt{2018MaYo}, \citeyear{2021MaMi}, \citealt{2022McYo}).

Theoretical studies have predicted that these populations also differ in their initial structural and kinematic properties. 
Specifically, hydrodynamical and $N$-body simulations have shown that 2G stars  may form in a sub-system more spatially concentrated than the 1G system  (see e.g. \citealt{2008DeVe}, \citealt{2011Bekk}, \citealt{2017BeJe}, \citealt{2019CaDe}, \citealt{2021LaCa}, \citeyear{2022LaCa}, \citealt{2022YaCa}, \citeyear{2022YaRo}).

Studying the dynamics of multiple populations from their initial to present-day dynamical state is necessary to connect the formation phase and the observed properties of these populations, as well as to understand how the global dynamical properties of globular clusters are affected by the presence of multiple populations with different spatial and kinematic properties.

The effects of dynamical processes acting during a cluster's long-term evolution alter the properties set at the time of formation and gradually erase the dynamical differences between 1G and 2G stars (see e.g. \citealt{2013MaPe}, \citeyear{2016MaPe}, \citealt{2015HeGi}, \citealt{2019TiVe}, \citealt{2021VeHo}, \citealt{2021sollima}). 
Some clusters, however, are expected to retain some of these differences, and recent observational studies have revealed dynamical differences between 1G and 2G stars generally consistent with those predicted by theoretical studies (see e.g. \citealt{2007SoFe}, \citealt{2009BePi}, \citeyear{2015BeVe}, \citealt{2011LaBe}, \citealt{2016SiMi}, \citealt{2017CoHe}, \citealt{2018MiMa}, \citealt{2019DaCa}, \citealt{2020CoMi}, \citealt{2022LiBe}, \citealt{2023onorato}, \citealt{2024MeMi};
See \citealt{2023LeBa} for a study showing two Galactic clusters (NGC6101 and NGC3201) with a 1G more concentrated than the 2G, but see also  \citealt{2024MeMi} for a recent study showing that in NGC3201 the 2G is more centrally concentrated than the 1G, and \citealt{2024CaDa} for an observational study of the dynamical properties of this cluster also providing support to scenarios in which the 2G formed more centrally concentrated than the 1G).

The dynamical differences between 1G and 2G stars imprinted by the formation processes and revealed by observations broaden the range of questions raised by the discovery of multiple populations to include a variety of issues concerning the evolution of the structural and kinematic properties of 1G and 2G stars.

The goal of the study presented in this paper is to explore the implications of initial differences between the spatial distributions of 1G and 2G stars for their evolution towards energy equipartition.
The evolution towards energy equipartition is one of the dynamical consequences of the collisional evolution of a stellar system (see e.g. \citealt{1987Spit}, \citealt{2003HeHu}).
As shown in a number of studies, the degree of energy equipartition in  globular clusters may provide a number of a key insights into their initial and present-day dynamical properties and stellar content such as, for example, the presence of stellar and intermediate-mass black holes, the anisotropy in the velocity distribution, the dynamical phase, and the core-collapsed nature of clusters (\citealt{1978Vish}, \citealt{2013TrvdM}, \citealt{2017WeVe}, \citealt{2018BiWeSi}, \citealt{2021PaVe}, \citeyear{2022PaVe}, \citealt{2022LiVe}, \citealt{2023ArVe}).
Thanks to high-precision HST proper motion studies spanning a broad range of stellar masses, the observational investigation of this aspect of the dynamics of stars clusters is now becoming possible (see e.g. \citealt{2018BeLi}, \citealt{2018LiBe}, \citeyear{2019LiBe}, \citeyear{2022LiBe}, \citealt{2022Watkins}).

Here we study the evolution towards energy equipartition in multiple-population globular clusters  through a suite of Monte Carlo simulations and explore how the degree of energy equipartition for different populations depends on time, distance from the cluster's center, and the velocity components.

This paper is organized as follows: we describe our simulations and the stars selected in our analysis in Section \ref{chap6:Methods}, in Section \ref{MP:SnKM} we present a brief general overview of the spatial and kinematic properties of the systems we have studied, our results are described in Section \ref{chap6:Results}, and we summarize our conclusions in Section \ref{chap6:Conclusions}.

\section{Initial conditions and Simulation Properties}
\label{chap6:Methods}
This study analyzes the dynamical evolution of five different models ran with the Monte Carlo simulation code \textsc{MOCCA} (\citealt{2013HyGi}, \citealt{2013GiHe}) on the Indiana University's Quartz supercomputer. The MOCCA code implements Henon's Monte Carlo method to follow the evolution of star clusters (\citealt{1971Heno}). \textsc{MOCCA} includes the effects of  two-body relaxation,  binary-binary and binary-single interactions (using the \textsc{FEWBODY} code by \citealt{2004FrCh}), and a spatial truncation mimicking the effects of the external tidal field of the host galaxy. In the simulations presented in this paper, binary and single stellar evolution are modeled using respectively, the BSE and SSE codes by \cite{2000HuPo,2002HuTo}. 
Supernovae kick velocities follow a Maxwellian distribution with a dispersion of 265 km/s (\citealt{2005HoLo}).
In a separate study we will investigate models with reduced kick velocities and higher retention fraction of black holes (see \citealt{2023ArVe} for a study of the role of stellar and intermediate-mass black holes on the evolution towards energy equipartition in single-population black holes).
For further details about the \textsc{MOCCA} code see \cite{2013HyGi}, \cite{2013GiHe} (see also \citealt{2022HyGi} for a description of recent updates to the code).

Our models include systems with a range of different values for the initial number of stars ($5\times10^5-2\times10^6$) following a \cite{KIMF} initial stellar mass function from $0.1-100$ $M_{\sun}$ with no primordial binaries; the effect of primordial binaries and the various parameters defining the primordial binary population will be addressed in a future paper. A metallicity of $Z=10^{-3}$ is adopted.
Each model has an initial ratio of the half-mass radius to the tidal cut-off radius  ($r_{\rm h}/r_{\rm t}$) of 0.14.
Each simulation includes two populations, 1G and 2G, initially described by King models \citep{1966King} with $W_{\rm 0,1G}=5$ for the 1G stars and $W_{\rm 0,2G}=7$ for the 2G stars, where the ratio of the initial half-mass radii of the 2G population to that of the 1G population ($r_{\rm h,1G}/r_{\rm h,2G}$) is equal to 20, and the initial fraction of 2G stars (defined as the ratio of the number of 2G stars to the total number of stars) is equal to 0.2. 
The system is initially set up in equilibrium with initial stellar velocities assigned using a velocity dispersion calculated from the Jeans equations (see e.g. \citealt{2008BiTr}) with the combined potential determined by the two stellar populations.
The choice of a 2G initially more centrally concentrated than the 1G is generally informed and motivated by the results of a number of theoretical studies predicting that 2G stars formed in the central regions of a more extended 1G system (see e.g. \citealt{2008DeVe}, \citealt{2010Bekk}, \citeyear{2011Bekk}, \citealt{2019CaDe}, \citealt{2022LaCa}).
For all models, the 1G initially extends to the cluster's truncation radius, and the truncation radii chosen are equal to tidal radii of clusters on circular orbits in a logarithmic potential at a galactocentric distance of 4 kpc (and 8 kpc for the model \SIMIV\ evolving in a weaker tidal field).
All the models start with an isotropic velocity distribution except for the model denoted by \SIMV\ which has structural properties identical to those of the \SIMII\ model but a radially anisotropic velocity distribution following an Osipkov-Merritt profile [$\beta = 1-(\sigma_{\theta}^2+\sigma_{\phi}^2)/(2\sigma_{\rm r})=1/(1+r^2_{\rm a}/r^2)]$, where $\sigma_{\theta}$, $\sigma_{\phi}$, and $\sigma_{\rm r}$ are the three spherical components of the velocity dispersion, and $r_{\rm a}$ is the anisotropy radius  (see e.g. \citealt{1979Osip}, \citealt{1985Merr}) with $r_{\rm a} = r_{\rm h}/2$.
The initial conditions are summarized in Table \ref{MPSimTable}.

We point out that the selection of models is not meant to provide a comprehensive exploration of the initial parameter space but rather to include a few selected choices aimed at illustrating the role of some of the key parameters that determine the rate at which two-body relaxation, stellar evolution, the response of the cluster to mass loss due to stellar evolution, and the strength of the external tidal field drive the clusters' dynamical evolution. 

For the following analysis, we focus on the main sequence stars of mass range $0.2-0.9$ $M_{\sun}$, and calculate the profiles and quantities found using the projected radius ($R$), projected radial velocity ($v_{\rm R}$), and projected tangential velocity ($v_{\rm T}$). 
The mass range chosen is similar to that approached in a number of recent studies of proper motion data (see e.g. \citealt{2022LiBe}, \citealt{2022Watkins}) and the high-end of the mass range roughly corresponds to the main sequence turn-off mass at 12 Gyr.

We perform the following analysis of projected radial profiles and time evolution using the average of 30 randomized projections for each snapshot, where one snapshot is taken each Gyr from 1 to 12 Gyr. Shaded regions will be provided for each plots, which represent the 25th to 75th percentiles of values from the 30 randomized projections.

\begin{table}
\centering
\begin{tabular}{|l|l|l|}
\hline
ID.  & $N$ & $r_{\rm t}$ \\ \hline
\SIMI\ & $2\times10^6$ & 97 pc   \\ 
\SIMII\   & $1\times10^6$ & 77 pc \\ 
\SIMV\   & $1\times10^6$ & 77 pc \\ 
\SIMIII\ & $5\times10^5$ & 61 pc  \\ 
\SIMIV\ & $5\times10^5$ & 97 pc  \\ 
\hline
\end{tabular}

\caption[Multiple-population simulations]{Summary of initial conditions for the \textsc{MOCCA} simulations presented in this study. 
All models in this study start with the same tidal filling factor ($r_{\rm h}/r_{\rm t}=0.14$), fraction of 2G stars ($N_{\rm 2G}/N_{\rm 1G+2G}=0.2$), $W_0$ parameters for the 1G and 2G ($W_{\rm 0,1G}=5$, $W_{\rm 0,2G}=7$), and ratio of half-mass radii of the 1G and 2G ($r_{\rm h,1G}/r_{\rm h,2G}=20$). \SIMV\ includes a radially anisotropic distribution with anisotropy radius $r_{\rm a}=r_{\rm h}/2$ (see Section \ref{chap6:Methods} for details). }
\label{MPSimTable}
\end{table}

\begin{figure}
    \centering
    \includegraphics[width=0.49\textwidth]{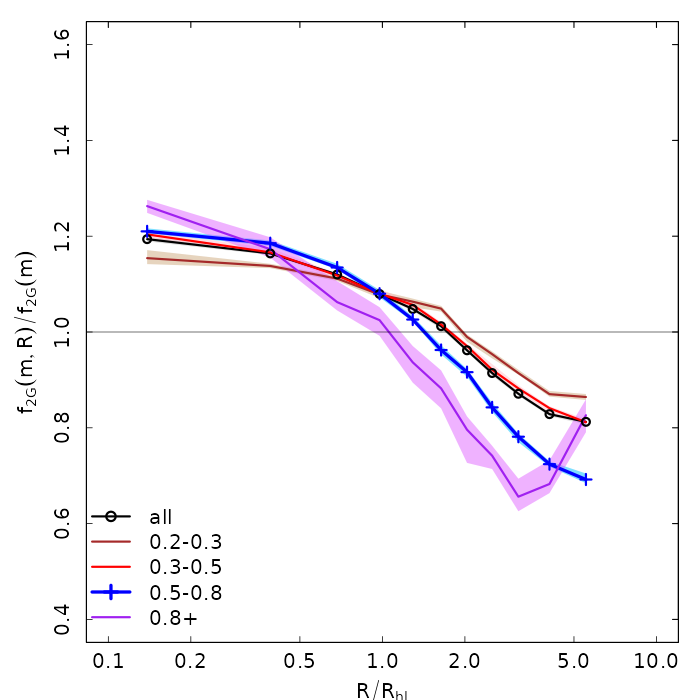}
    \caption[Radial Mass Mixing]{Radial profile of the fraction of 2G stars by stellar mass group (as defined in the legend) normalized by the global fraction of 2G stars in the same stellar mass group, of \SIMII\ at $t=12$ Gyr. 
    The projected radius ($R$) is normalized by the projected half-light radius of the cluster ($R_{\rm hl}$).
    The degree of mixing slightly increases for decreasing values of the stellar masses.}
    \label{fig:RadialMassmixing}
\end{figure}
\begin{figure}
    \centering
    \includegraphics[width=0.49\textwidth]{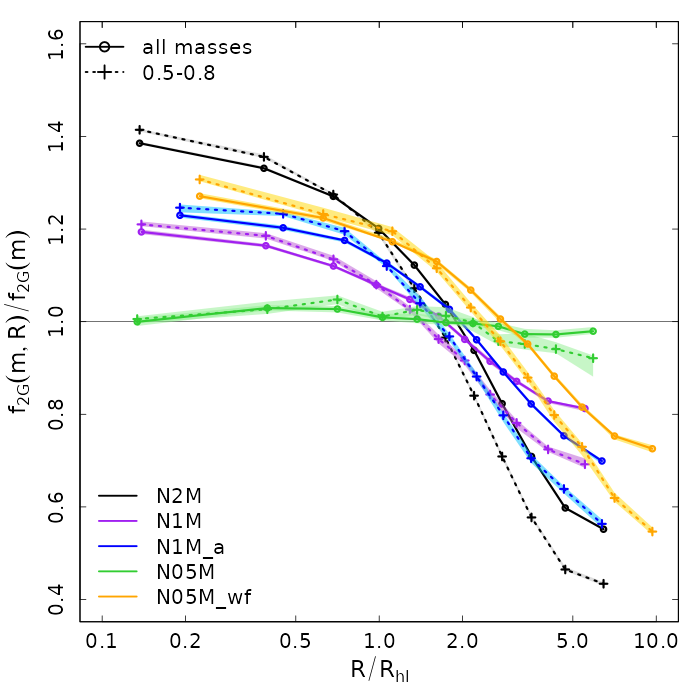}
    \caption[Mass Mixing Synthesis]{Radial profile of the fraction of 2G stars for two stellar mass groups (as defined in the legend), normalized by the global fraction of 2G stars by stellar mass group, at $t=12$ Gyr for each model presented in this study.}
    \label{fig:MassMixSynth}
\end{figure}
\begin{figure*}
    \centering
    \includegraphics[width=\textwidth]{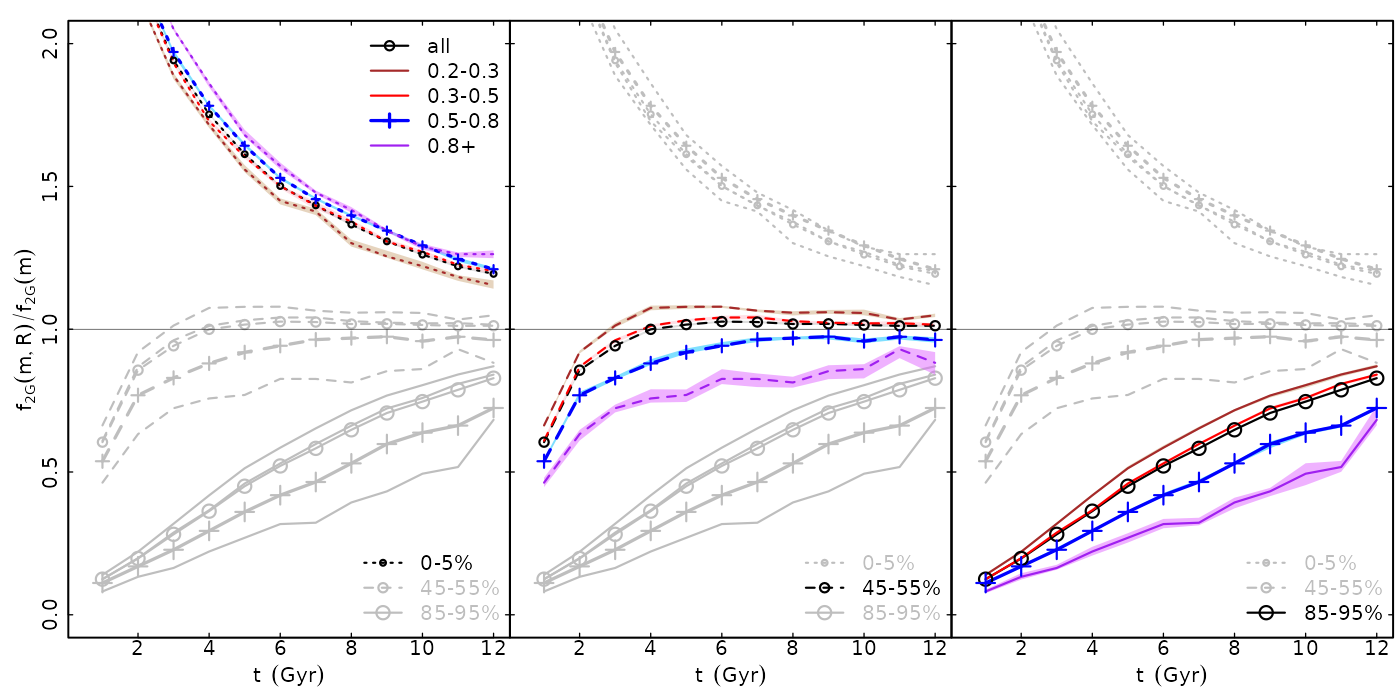}
    \caption[Time Evolution of Mass Mixing]{Time evolution of the ratio defined in Figure \ref{fig:RadialMassmixing}, for three Lagrangian radial shells of \SIMII. Each panel shows the same set of lines but highlights the time evolution of one radial shell. Complete mixing would correspond to the case in which the value of the ratio plotted is equal to one for all mass groups and at all radial distances. The local value of the fraction of second generation stars measured close to the half-mass radius is similar to the global value for most of the cluster's evolution.}
    \label{fig:TimeMassmixing}
\end{figure*}
\begin{figure}
    \centering
    \includegraphics[width=0.49\textwidth]{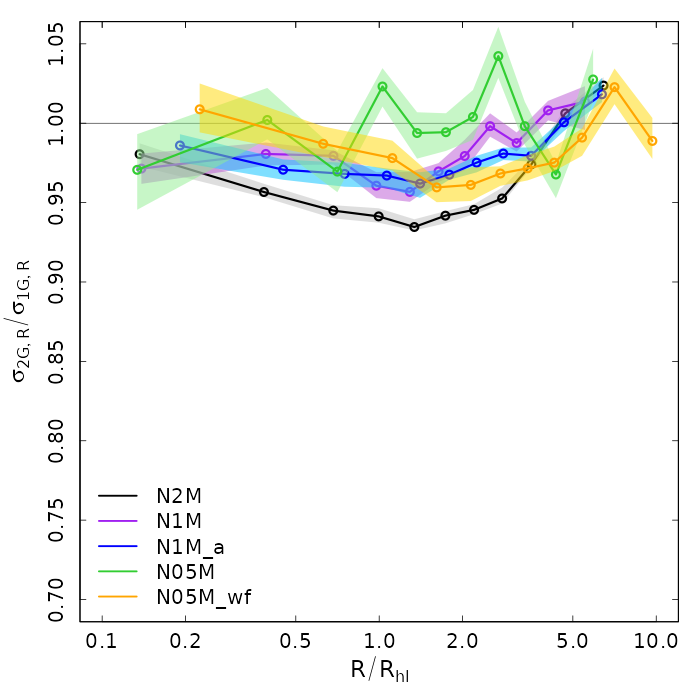}
    \includegraphics[width=0.49\textwidth]
    {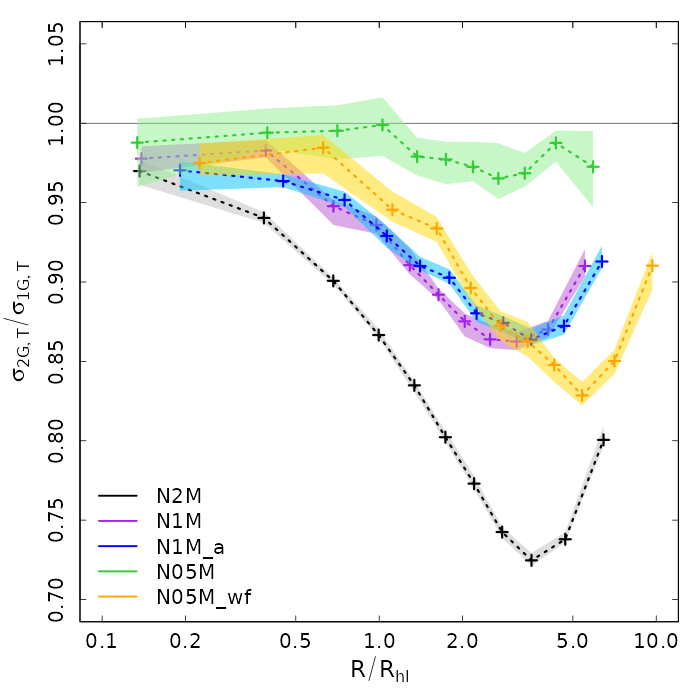}
    \caption[Kinematic Mixing Synthesis]{Radial profile of the ratio of the velocity dispersion of the 2G stars to 1G stars for the radial (top) and tangential velocity components (bottom) at $t=12$ Gyr across all models in this study.}
    \label{fig:DispMixSynth}
\end{figure}
\begin{figure*}
    \centering
    \includegraphics[width=\textwidth]{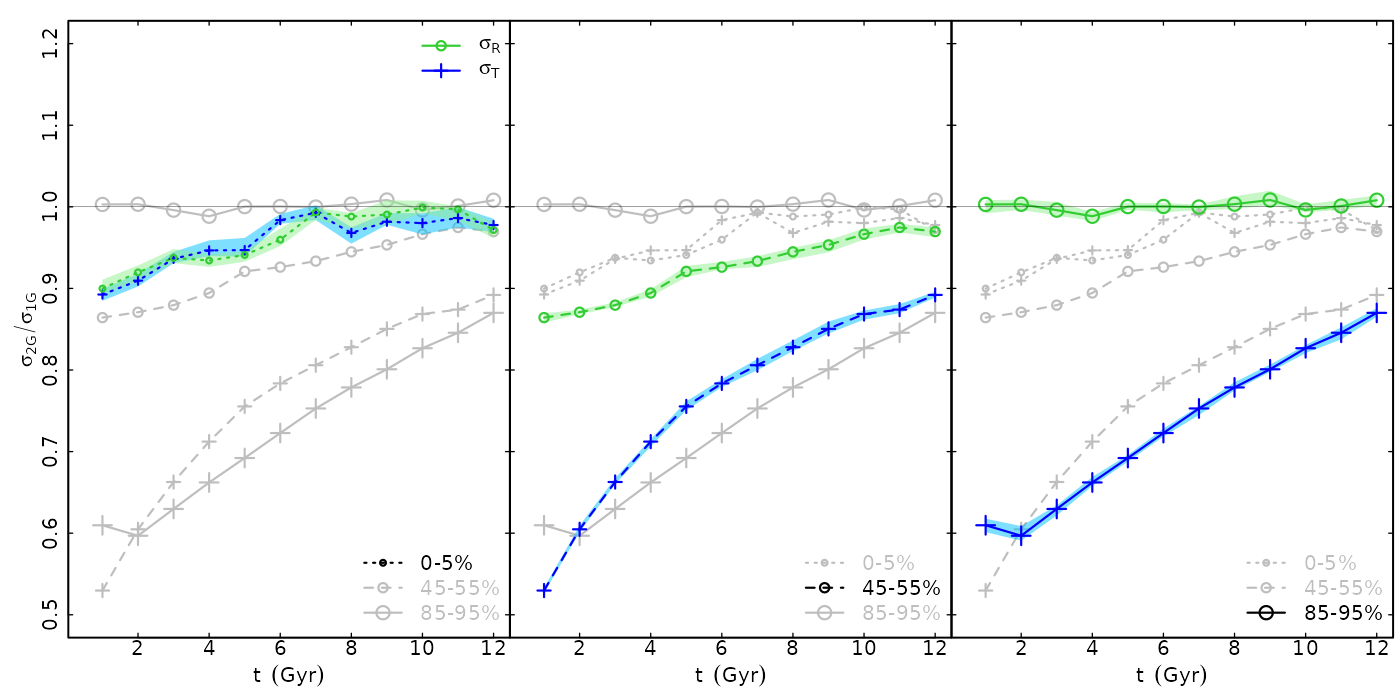}
    \caption[Time Kinematic Mixing]{Time evolution of the ratio of the velocity dispersion of the 2G stars to 1G stars for the radial and  tangential velocity components in \SIMII\ in multiple radial shells. Each panel shows the same set of lines but highlights the time evolution of one radial shell. Except for the innermost regions, the tangential velocity dispersion of the 2G and 1G stars is characterized by significant differences for most of the cluster evolution.}
    \label{fig:TimeKinMix}
\end{figure*}
\begin{figure*}
    \centering
    \includegraphics[width=\textwidth]{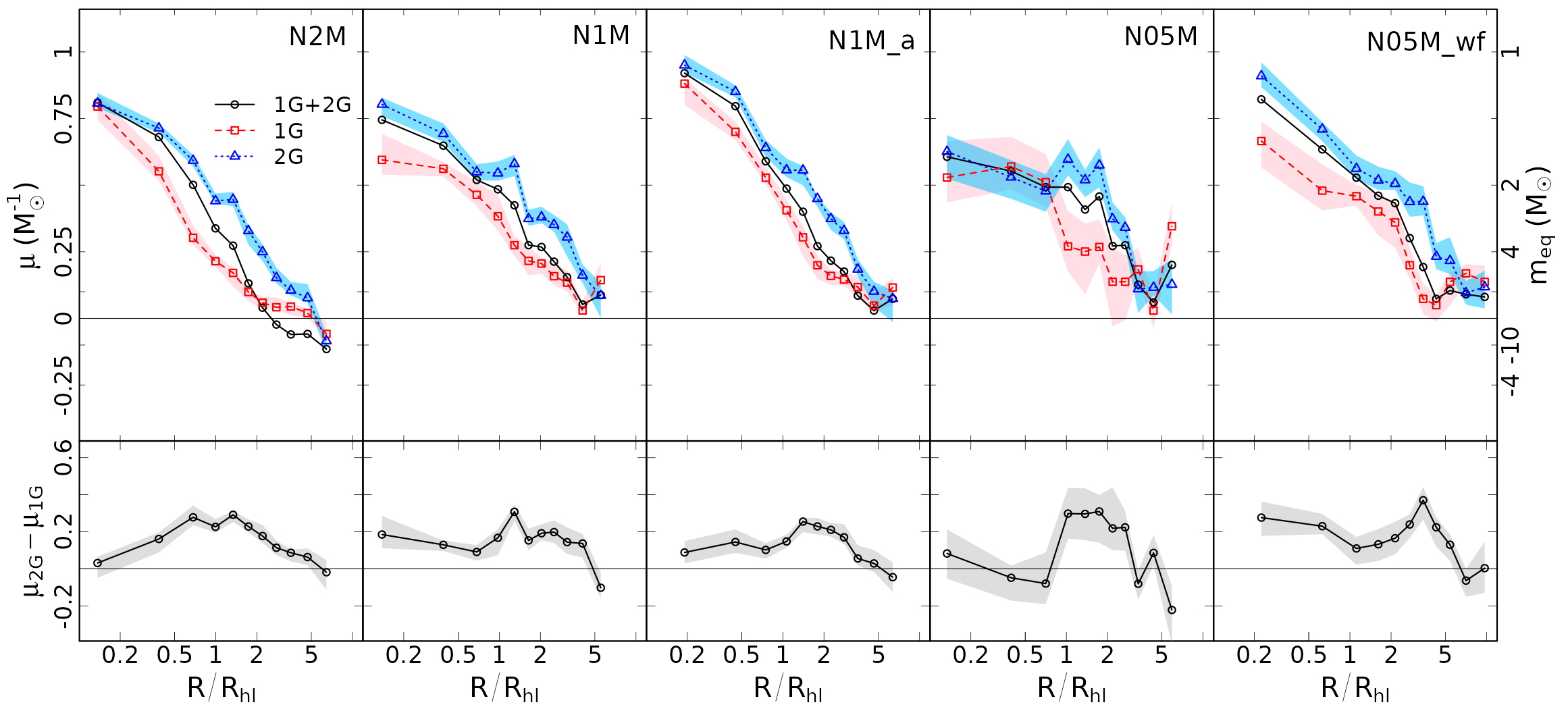}
    \caption[Radial Equipartition Profile]{Radial profile of the inverse of the equipartition mass, $\mu$ (and equipartition mass, $m_{\rm eq}$, on the secondary $y$-axis), calculated using both velocity components for all models at $t=12$ Gyr. The bottom sub-panels show the radial profile of the difference in $\mu$ between 2G and 1G stars. The degree of energy equipartition increases at smaller distances from the cluster's center. In general,  the 2G population is characterized by a stronger degree of energy equipartition than the 1G population.}
    \label{fig:meq_prof_tot}
\end{figure*}
\begin{figure*}
    \centering
    \includegraphics[width=\textwidth]{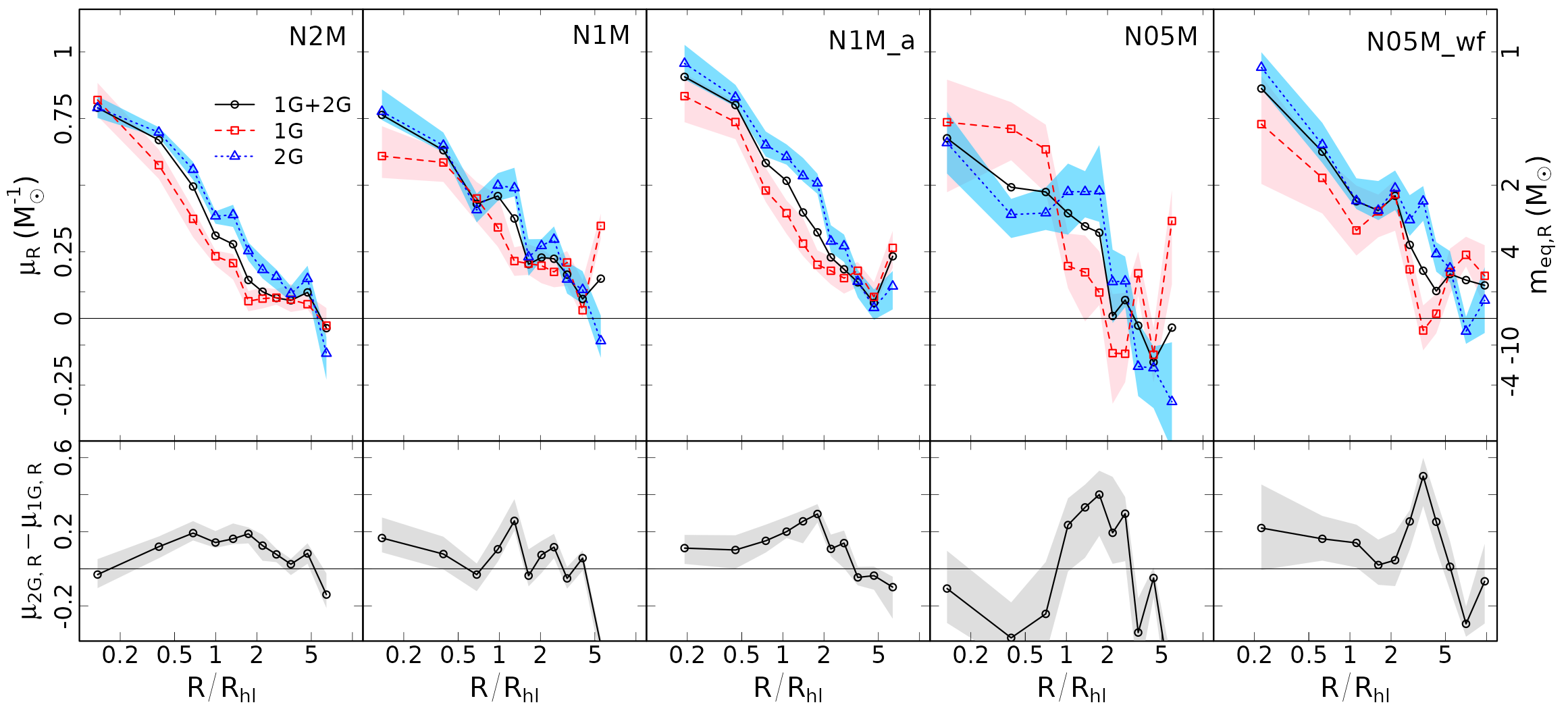}
    \includegraphics[width=\textwidth]{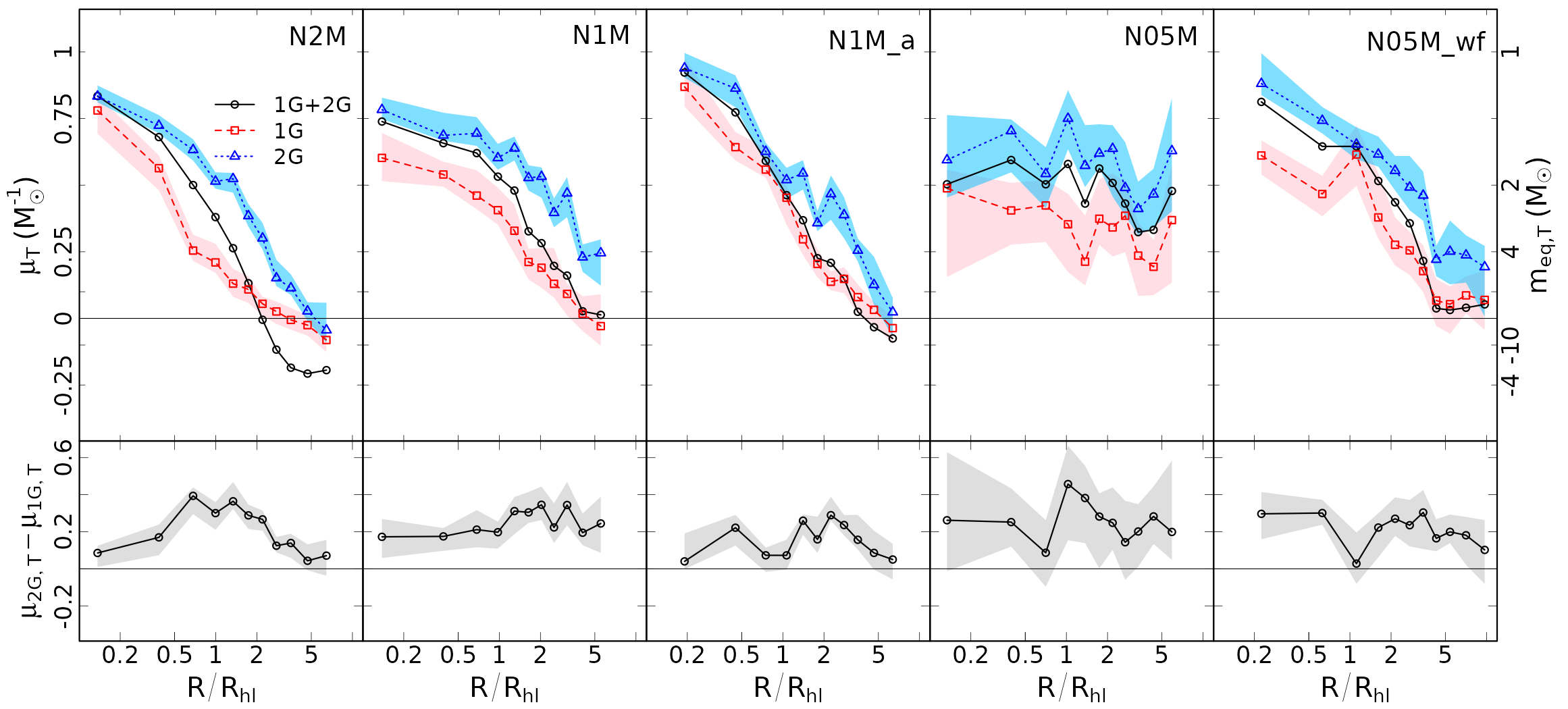}
    \caption[Radial Equipartition Profile]{Same as Figure \ref{fig:meq_prof_tot}, calculated using the radial ($\mu_{\rm R}$, top) and tangential ($\mu_{\rm T}$, bottom) velocity components for all models at $t=12$ Gyr. The bottom sub-panels of each panel show the radial profile of the difference in $\mu$ between 2G and 1G stars. The differences between the two generations are more significant for the degree of energy equipartition calculated using the tangential component of the velocity dispersion. Note that the radial and tangential inverse equipartition mass show different radial profiles within each simulation.}
    \label{fig:meq_prof}
\end{figure*}

\begin{figure}
    \centering
    \includegraphics[width=0.4\textwidth]{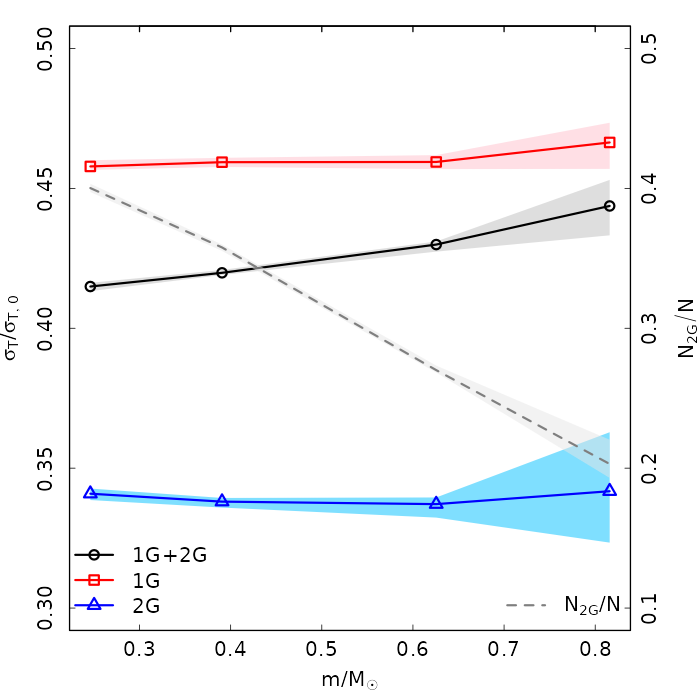}
    \caption[Radial Equipartition Profile]{Tangential velocity dispersion, normalized by the central velocity dispersion, versus mass for the \SIMI\ model for each population and both populations combined for the 85\%$-$95\% radial regions at $t=12$ Gyr. On the secondary $y$-axis, we show the fraction of 2G stars in each mass bin in the same radial region. For both the 2G and 1G stars, the velocity dispersion does not depend on the stellar mass; the inverted trend of velocity dispersion increasing with stellar mass when all stars are considered together results from both the variation of the fraction of 2G stars with the stellar mass and the difference in velocity dispersion between the 2G and 1G stars.}
    \label{fig:meq_unmixed}
\end{figure}

\begin{figure*}
    \centering
    \includegraphics[width=\textwidth]{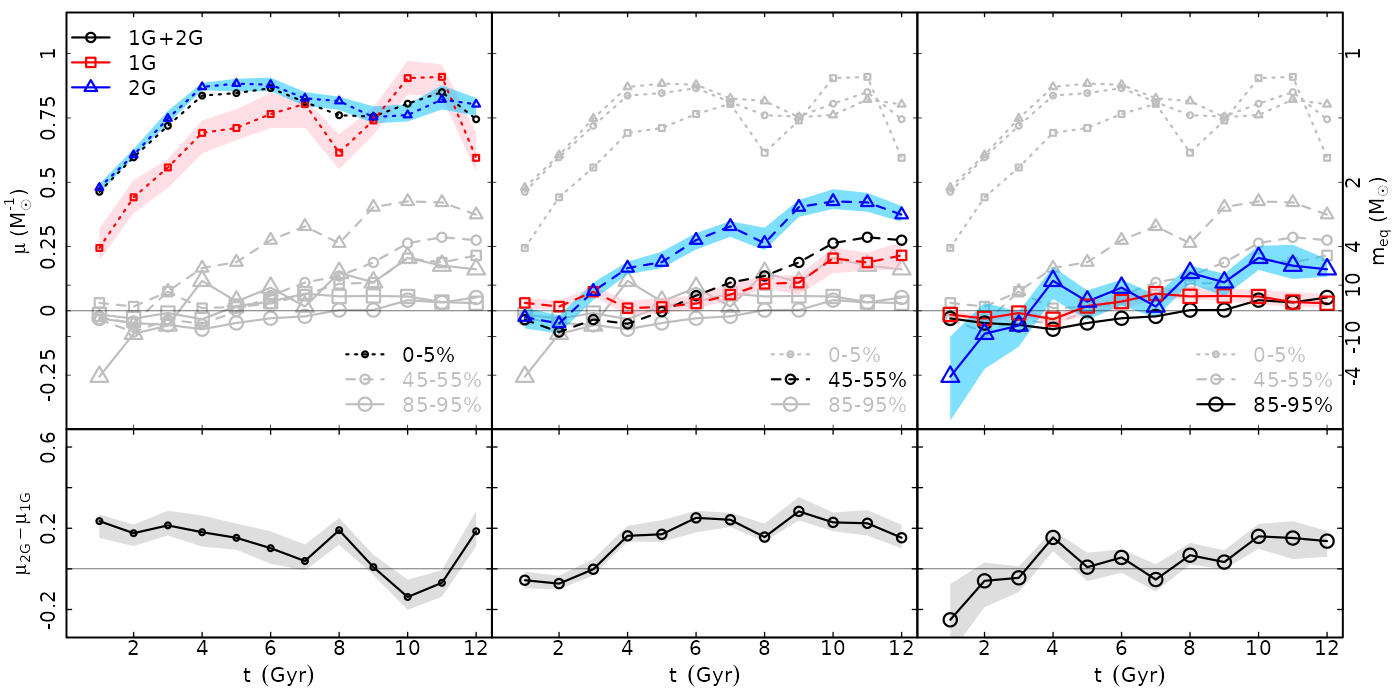}
    \caption[Equipartition Time Evolution for Total Dispersion]{Time evolution of the inverse of the equipartition mass, $\mu$, (and equipartition mass, $m_{\rm eq}$, on the secondary $y$-axis)  calculated using the total velocity dispersion in multiple radial shells for \SIMII. Each panel shows the same set of lines but highlights the time evolution of one radial shell. The bottom sub-panel of the plot shows the difference between the two populations as a function of time. The inner regions are those characterized by the most rapid evolution towards energy equipartition and at all radii the evolution is more rapid for the 2G population.}
    \label{fig:meq_evo_tot}
\end{figure*}
\begin{figure*}
    \centering
    \includegraphics[width=\textwidth]{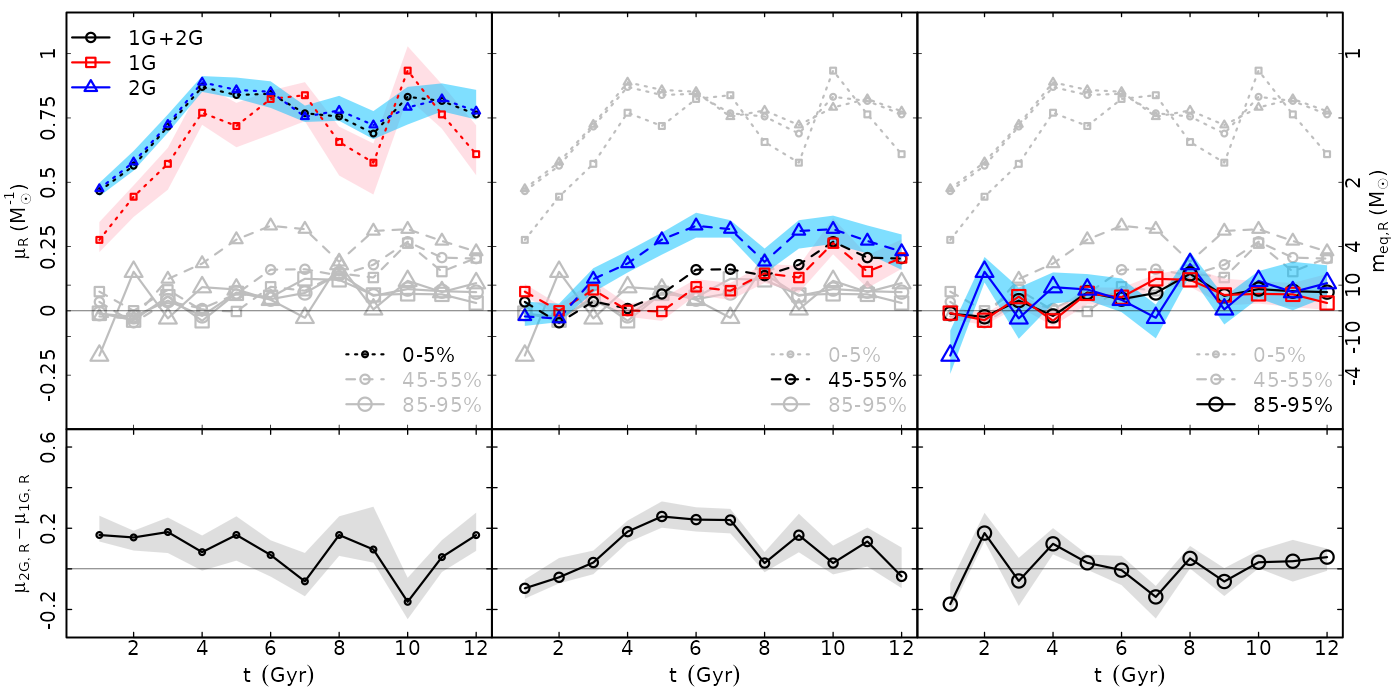}
    \includegraphics[width=\textwidth]{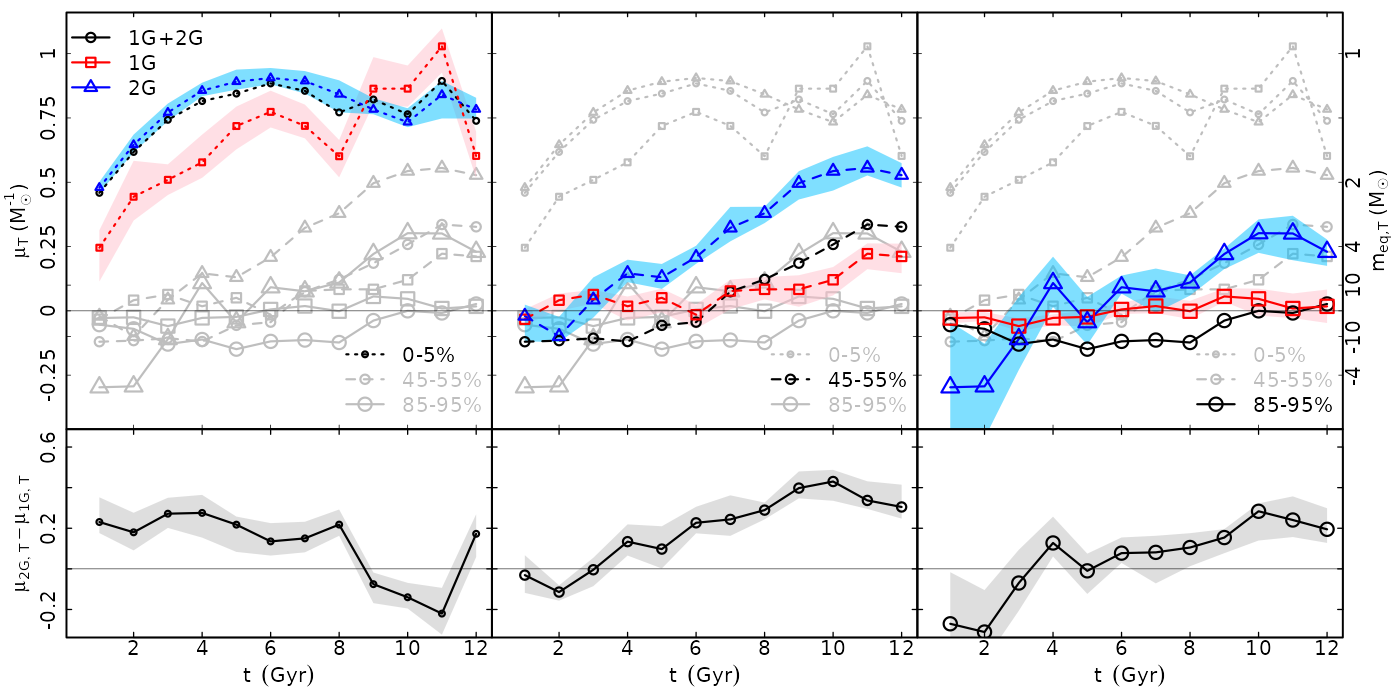}
    \caption[Equipartition Time Evolution for Each Component]{Same as Figure \ref{fig:meq_evo_tot}, for the equipartition mass as evaluated from the radial (top panels) and the tangential (bottom panels) velocity components. The bottom region of the plots show the difference between the two populations as a function of time. The equipartition mass in the tangential velocity component shown in the bottom panels displays the strongest differences between the 1G and the 2G stars.}
    \label{fig:meq_evo_Rphi}
\end{figure*}
\begin{figure*}
    \centering
    \includegraphics[width=\textwidth]{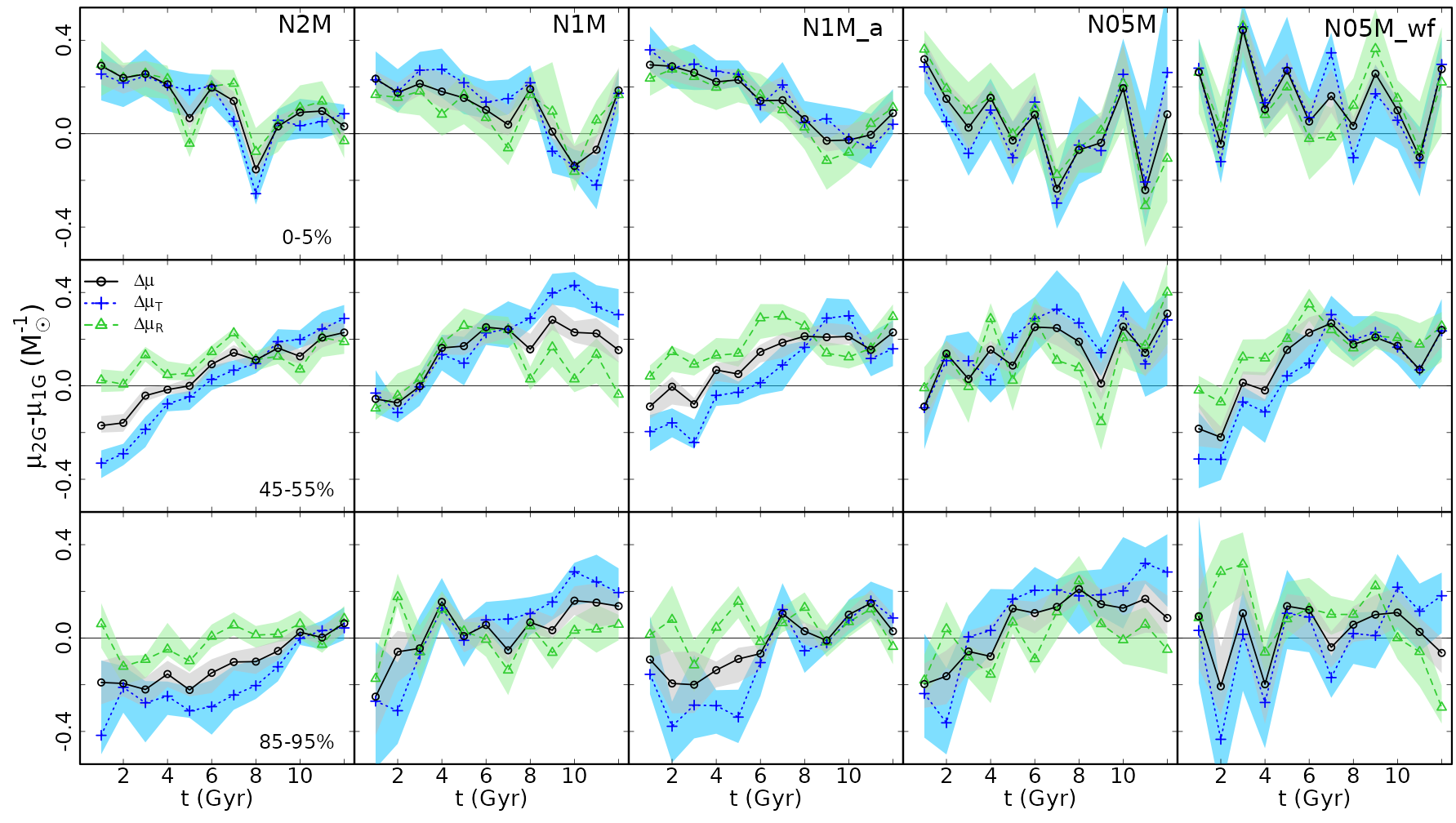}
    \caption[Equipartition Time Evolution for Total Dispersion]{Time evolution of the difference of the inverse equipartition mass between the 2G and 1G populations, $\Delta\mu=\mu_{\rm 2G}-\mu_{\rm 1G}$, for different radial shells (rows) and each simulation (columns). Each panel shows the difference in the total, tangential, and radial components of the equipartition. The intermediate regions show the difference flip from negative to positive, and are strongest of the radial regions shown at 12 Gyr.}
    \label{fig:meq_diff_evo}
\end{figure*}

\begin{figure*}
    \centering
    \includegraphics[width=\textwidth]{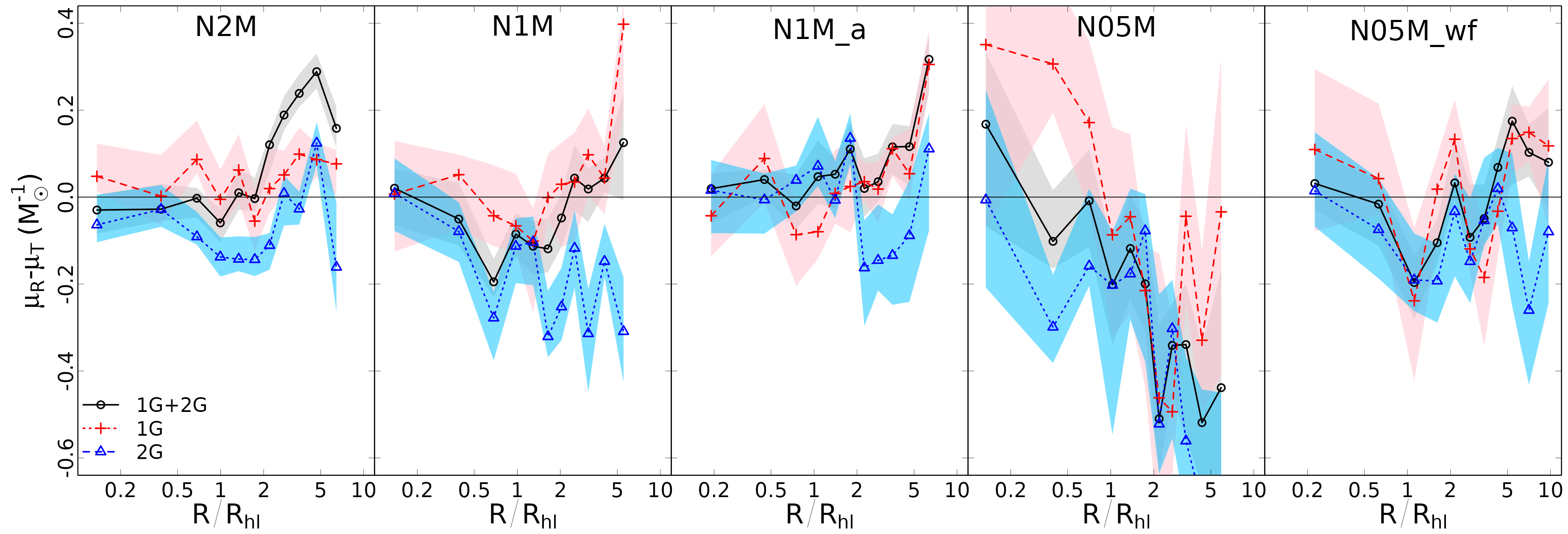}
    \caption[Anisotropy in Mu]{Radial profile of the difference between the radial and tangential inverse equipartition masses, $\mu_{\rm R}-\mu_{\rm T}$, for each population and in each model at 12 Gyr. Note that the systems are "isotropic" in energy equipartition in the inner regions, and for some models are anisotropic in the intermediate to the outer regions.}
    \label{fig:ani_prof}
\end{figure*}

\begin{figure*}
    \centering
    \includegraphics[width=\textwidth]{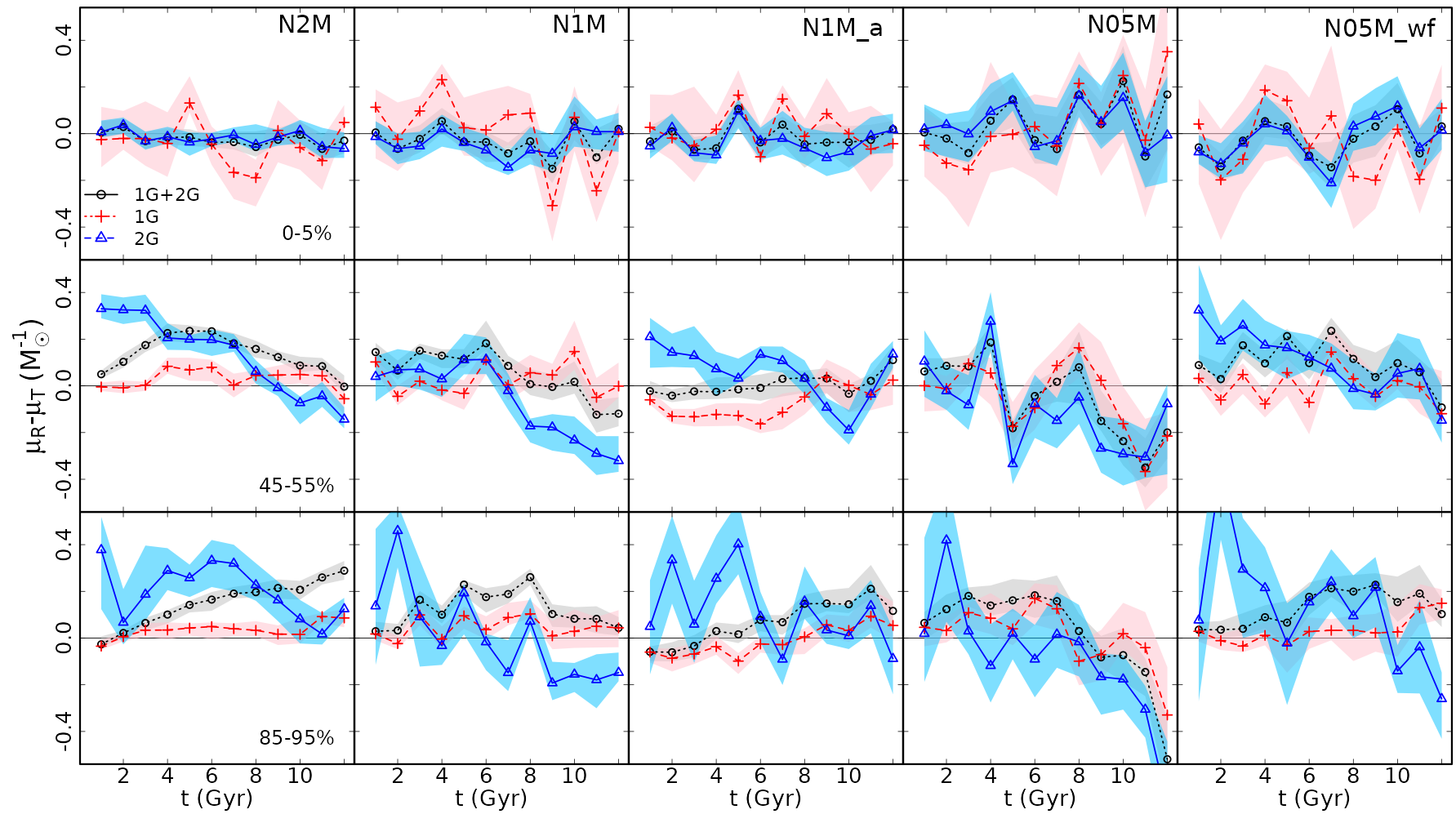}
    \caption[Anisotropy in Mu]{Time evolution of the difference between the radial and tangential inverse equipartition masses, $\mu_{\rm R}-\mu_{\rm T}$, for each population and a range of radial shells in each model.}
    \label{fig:meqani_evo_tot}
\end{figure*}

\section{An overview of the spatial and kinematic properties}
\label{MP:SnKM}
The focus of this paper is on the evolution towards energy equipartition in 1G and 2G stars in multiple-population globular clusters. Before discussing the results concerning energy equipartition, we start with a brief preliminary overview of the spatial and kinematic properties of the systems investigated and the degree of dynamical mixing of the 1G and 2G properties.

\subsection{Spatial Mixing}

In Figure \ref{fig:RadialMassmixing}, we plot the fraction of 2G stars for multiple mass ranges, $f_{2G}(m,R)$, normalized by the global 2G fraction of that mass range, $f_{2G}(m)$, as a function of projected radius (normalized by the projected half-light radius, $R_{\rm hl}$) for the \SIMII\ model at 12 Gyr. 
Note that a completely mixed cluster would be represented as a horizontal line at  $f_{2G}(m,R)$/$f_{2G}(m)=1$ for all mass bins.
By 12 Gyr, \SIMII\ exhibits a partial level of radial mixing across all stars and high-mass stars are further from mixed than low mass stars.

We find a range of similar trends across all models in Figure \ref{fig:MassMixSynth}, where we plot the mixing of all stars selected and of a high-mass bin ($0.5-0.8\ M_{\sun}$; dashed line) for all simulations. 
For reference, mean mass of the stars in our selection and in our total cluster at 12 Gyr are $\sim0.42-0.47\ M_{\sun}$ and $\sim$ 0.39 $M_{\sun}$, respectively, except for the \SIMIII\ model, which has an average stellar mass of the total cluster of $\sim$ 0.49 $M_{\sun}$. 
\SIMIII\ is the system with the shortest initial half-mass relaxation time and nearly completely mixed by 12 Gyr, while \SIMI\ is the dynamically youngest and the furthest model from complete spatial mixing.

It is interesting to notice the differences between the degree of mixing of the \SIMII\ and \SIMV\ models; these two systems have initially the same structural properties but differ in their initial velocity anisotropy radial profile (see Section \ref{chap6:Methods}); although the differences in the degree of spatial mixing of these two models is modest, it suggests differences in the initial kinematic properties may play a role in the rate of spatial mixing. We will further explore the role of the initial anisotropy on the spatial mixing rate in a future investigation.

\begin{table}
\centering
\begin{tabular}{|l|l|l|l|}
\hline
ID.  & $N$ & $R_{\rm hl}(pc)$ & $M_{\rm 2G}/M_{\rm tot}$\\ \hline
\SIMI\ &539,702 & 3.73  &  0.58 \\ 
\SIMII\   & 200,989 &3.23 & 0.60 \\ 
\SIMV\   & 259,938 &3.01 &0.52 \\ 
\SIMIII\ & 43,864 & 1.98 &0.61  \\ 
\SIMIV\ & 116,719 & 2.50 &0.57  \\ 
\hline
\end{tabular}

\caption[Multiple-population simulations at 12 Gyr]{Values at 12 Gyr of the total number of stars, $N$, half-light radius, $R_{\rm hl}$, and ratio of the total mass is 2G stars to the total cluster mass, $M_{\rm 2G}/M_{\rm tot}$.}
\label{12GyrTable}
\end{table}

To further illustrate the dynamical evolution of the radial mixing, in Figure \ref{fig:TimeMassmixing} we show the mixing ratio shown in Figure \ref{fig:RadialMassmixing} as a function of time for multiple Lagrangian projected-radius bins (calculated using all stars selected), and shown in units of the projected half-light radius in Table \ref{Rtab}) for a representative model, \SIMII.
This figure clearly shows the development of a mass-dependent mixing during the cluster long-term evolution; this behavior matches that previously found in \cite{2021VeHo} but, as already pointed by those authors, the dependence of the degree of mixing on the stellar mass may be too weak to be detected in observational data. 
Finally, in Table \ref{12GyrTable}, we report the values of the total number of stars, the half-light radius, and the global fraction of the total cluster mass in 2G stars at $t=12$ Gyr.
The limited set of initial conditions considered for this initial exploration of energy equipartition in multiple-population clusters are not meant to produce systems spanning the entire range of observed clusters' properties, 
but the final properties of our models reported in Table \ref{12GyrTable} are generally consistent with those typically found in many Galactic globular clusters\footnote{
See e.g. \citealt{2017MiPi} for an extensive study of multiple populations in Galactic globular clusters based on HST observation. They show that the distribution of the fraction of 2G stars in their sample has a median of 64\% with the 25th and 75th percentiles equal to 55\% and 72\% respectively. The full range of 2G's fraction in their sample was 35\% to 90\%.}
(see also \citealt{2021VeHo}, \citealt{2024HyVe} for more extensive investigations studying the dependence of the 2G fraction on the clusters’ initial conditions). 
We also note that Figure \ref{fig:MassMixSynth} shows that our models span various degree of spatial mixing at 12 Gyr encompassing the variety of spatial distributions of 1G and 2G stars found in Galactic globular clusters ranging from clusters where the two populations are completely mixed (see e.g. \citealt{2014DaMa}, \citealt{2015NaMi}) to those in which some memory of the initial differences is still present and the 2G population is more centrally concentrated than the 1G population (see e.g. \citealt{2007SoFe}, \citealt{2009BePi}, \citealt{2011LaBe}, \citealt{2012MiPi}, \citealt{2014CoPi}, \citealt{2016SiMi}, \citealt{2019DaCa}, \citealt{2023onorato}). 
Similar initial conditions (see e.g. \citealt{2021VeHo}) have also been shown to produce differences in the anisotropy between 1G and 2G stars generally consistent with those found in the Galactic clusters for which the kinematics of multiple populations has been studied (see e.g. \citealt{2015BeVe}, \citealt{2018MiMa}, \citealt{2020CoMi}, \citeyear{2024CoCa}, \citealt{2024DaCa}; an extensive investigation of the spatio-kinematical mixing will be presented in Aros et al. in prep.).

\subsection{Kinematic mixing}
We continue our analysis with a brief overview of our results concerning the kinematic mixing of the two populations in the projected radial and tangential directions.

In Figure \ref{fig:DispMixSynth}, we show the radial profile of the ratio of the 2G to the 1G velocity dispersion for the radial (top panel) and tangential (bottom panel) velocity components measured at 12 Gyr for all the models. 
We point out that while the two populations are close to kinematically mixed in the radial component of the velocity dispersion, there are significant differences between the tangential velocity dispersions of the two populations, especially in the intermediate regions.
This result is consistent with the predictions of previous numerical studies (see e.g. \citealt{2015BeVe}, \citealt{2021VeHo}) and with the findings of the first observational investigations of the kinematic properties of multiple stellar populations (see e.g. \citealt{2015BeVe}, \citeyear{2018BeLi}, \citealt{2020CoMi}, \citealt{2022LiBe}).

In order to illustrate the evolutionary path leading to the final radial profiles shown in Figure \ref{fig:DispMixSynth}, we plot the time evolution of those velocity dispersion ratios for a selection of radial shells in Figure \ref{fig:TimeKinMix} for the \SIMII\ model. See Table \ref{Rtab} for the radii of each selected shell at 12 Gyr.

\begin{table}
\centering
\begin{tabular}{|l|l|l|l|l|l|l|}
\hline
ID.  & $R_{5\%}$ & $R_{45\%}$ & $R_{55\%}$ & $R_{85\%}$ & $R_{95\%}$ \\ \hline
\SIMI\ & 0.23& 1.5& 2.0& 4.1&5.7     \\ 
\SIMII\    & 0.23& 1.5& 1.8& 3.5&4.9   \\ 
\SIMV\   & 0.29& 1.6& 2.0& 4.0&5.6 \\ 
\SIMIII\ & 0.23& 1.6& 2.0& 3.8&5.2 \\ 
\SIMIV\ & 0.37& 2.4& 3.1& 6.1&8.5  \\ 
\hline
\end{tabular}
\caption[Multiple-population simulations]{Summary of Lagrangian radial shells by number (in units of $R_{\rm hl}$) for the selections we use for all models in this study (main sequence stars with masses $0.2-0.9M_{\sun}$) at 12 Gyr.}
\label{Rtab}
\end{table}

This shows that as the 2G stars migrate towards the outer regions, they are characterized by a radial velocity dispersion which is close to that of the local 1G stars; the 2G tangential velocity dispersion, on the other hand, is smaller than that of 1G stars and evolves towards the 1G values at a rate dependent on the distance from the cluster's center. In the intermediate and outer regions (see, respectively, the  middle and right-hand panels of Figure \ref{fig:TimeKinMix}) the differences between the 2G and the 1G tangential velocity dispersions are still not negligible when the radial velocity components are already mixed.
These differences between the tangential velocity dispersions of the two stellar generations are responsible for the stronger radial anisotropy of the 2G population found in previous studies (see e.g. \citealt{2015BeVe}, \citealt{2019TiVe}, \citealt{2021VeHo}, \citealt{2023LibVe}).
A detailed investigation of the evolution of the 1G and 2G anisotropy and of the distribution of 1G and 2G stars in phase space will be presented in a future paper (Aros et al., in prep.).

\section{Energy equipartition}
\label{chap6:Results}

\subsection{Evolution towards energy equipartition}
We now turn our attention to the study of the evolution towards energy equipartition of the 1G and 2G populations and how this evolution is affected by the differences in the initial dynamical properties of the two populations.

In our analysis, we evaluate the degree of energy equipartition at a given distance from the cluster's center using the equipartition mass, $m_{\rm eq}$, (from \citealt{2022LiVe} and \citealt{2023ArVe}, modified from the original definition in \citealt{meq} to allow for negative equipartition masses). The equipartition mass is a function of the distance from the cluster's center, and at a given clustercentric distance is defined as follows:

\begin{equation}
    \sigma(m) =\left\{\begin{array}{cc}
    \sigma(m_{\rm eq}) \left(\frac{m}{m_{\rm eq}}\right)^{-1/2} &  \rm{if} \ m> m_{\rm eq}\ \ \rm{and} \ m_{\rm eq}>0 \\
    \sigma_0 \exp\left(-\frac{1}{2} \frac{m}{m_{\rm eq}}\right) & \rm{otherwise.}
         \end{array}  \right.
    \label{eq:meq4}
\end{equation}

\noindent where $\sigma$ is the expected velocity dispersion for a given mass, $\sigma_0$ is the limit of the velocity dispersion at mass 0, and $m_{\rm eq}$ is the equipartition mass, 
corresponding to the value of the stellar mass such that (for $m_{\rm eq}>0$) stars more massive than $m_{\rm eq}$ follow the complete  energy equipartition relation.
Note that the degree of energy equipartition changes with radius, where the central regions of globular clusters develop a higher degree of equipartition than the intermediate and outer regions (see, e.g. \citealt{2013TrvdM}, \citealt{2017WeVe}, \citealt{2021PaVe}, \citeyear{2022PaVe}, and \citealt{2023ArVe}; see also \citealt{2018LiBe}, \citealt{2022Watkins} for observational evidence).

We fit the equipartition mass with the radial velocities ($m_{\rm eq, R}$) and tangential velocities ($m_{\rm eq, T}$) individually through the following likelihood function:

\begin{equation}
\mathcal{L}=\displaystyle \prod_{i=1}^N\frac{1}{\sqrt{2\pi \sigma^2(m_{\rm i})}}\exp\left[ \frac{-v_{\rm i}^2}{2\sigma^2(m_{\rm i})} \right]
\end{equation}

\noindent where $m_{\rm i}$ and $v_{\rm i}$ are, respectively, the mass and velocity of each star at a given distance from the cluster's center.
Additionally, we calculate the equipartition mass from both the tangential and radial velocity components combined ($m_{\rm eq}$) using the following likelihood function:

\begin{equation}
\mathcal{L}=\displaystyle \prod_{i=1}^N\frac{1}{2\pi \sigma^2(m_{\rm i})}\exp\left[ \frac{-(v_{\rm R, i}^2+v_{\rm T, i}^2)}{2\sigma^2(m_{\rm i})} \right]
\label{eq:toteq}
\end{equation}

\noindent where in this case, $\sigma$ represents a singular 1D total velocity dispersion from both tangential and radial velocities. Note that we often use the inverse of the equipartition mass ($\mu$=1/$m_{\rm eq}$), where 0 indicates no equipartition, larger values indicate a stronger decline of velocity dispersion with higher masses, and negative values indicate an increase of the velocity dispersion with higher masses (which is a trend opposite to that of clusters evolving towards energy equipartition, see \citealt{2022PaVe} for an example of how clusters can evolve towards an inverted equipartition in the outer regions).

\subsubsection{Radial variation of the degree of energy equipartition}

In Figure \ref{fig:meq_prof_tot}, we plot the radial variation of the degree of equipartition (calculated using both components of the velocity dispersion, see equation \ref{eq:toteq}) for both populations combined and each population separately for all models at 12 Gyr. Additionally, this figure includes a plot of the radial variation of the difference between $\mu$ of the 2G and 1G populations.

We find that the largest differences between the degree of energy equipartition of 1G and 2G stars are typically found in the cluster's intermediate regions (at $1<R/R_{hl}<3$) with the 2G characterized by a stronger equipartition than the 1G; this trend agrees with the initial findings of \cite{2021VeHo}.
We interpret this trend as a consequence of the fact that the 2G stars currently in the outer regions formed in a denser inner sub-system where, as they were gradually diffused outwards, they underwent a more rapid evolution towards energy equipartition.
The difference between the degree of energy equipartition of 1G and 2G stars in the inner regions is generally smaller than that found in the intermediate regions while no significant differences are found in the outermost regions.

The radial variation of the degree of equipartition for all the stars (i.e. for the 1G and 2G stars selected, combined) depends on the radial variation of the fraction of 1G and 2G stars: in the inner regions, where the 2G population is dominant, it is closer to the degree of energy equipartition of the 2G stars, while in the outermost regions it approaches that of the 1G stars.

To analyze the role of each velocity component in the degree of energy equipartition, we evaluate separately in Figure \ref{fig:meq_prof} the radial component (top panel) and the tangential component (bottom panel) for each population individually and combined.
While we find the same radial trend and stronger level of energy equipartition in the 2G population as seen in Figure \ref{fig:meq_prof_tot}, we also find a stronger degree of energy equipartition in the tangential component than the radial component of the velocity dispersion. 
In both components, the 2G population is at a more advanced stage of its evolution towards energy equipartition, but the difference between the degrees of energy equipartition is larger in the tangential component as seen in the bottom sub-panels of each panel.

When looking at the \SIMII\ and \SIMV\ models, we find the radial profiles of energy equipartition are different between the models when using both velocity components or the tangential component alone. 
The difference between the populations is slightly stronger in the radial component in the \SIMV\ model than the \SIMII\ model, and vice-versa when using the tangential component. These results further show that for systems with similar initial structural properties, the development towards energy equipartition depends also on the level of anisotropy in the cluster (see e.g. \citealt{2021PaVe}, \citeyear{2022PaVe}, \citealt{2024PaHe}; see also \citealt{2022LiVe} for the dependence on the strength of the initial rotation).

It is interesting to note that in the \SIMIII\ model the degree of tangential equipartition of the 2G stars at 12 Gyr is still significantly different from that of the 1G stars despite the fact that, as shown in Section \ref{MP:SnKM}, the two populations are spatially (Figure \ref{fig:MassMixSynth}) and kinematically mixed (Figure \ref{fig:DispMixSynth}). 
This result suggests that energy equipartition may also reveal differences in the dynamical history of the two populations in dynamically old clusters, where the memory of other differences that were imprinted by formation and evolutionary processes have been lost.

\subsubsection{The effect of multiple, unidentified kinematic populations on the measure of the energy equipartition}

This section will focus on the possible impact of measuring the equipartition mass of the entire stellar content of a multiple-population cluster without first identifying the two populations.

As discussed above, the evolution towards energy equipartition proceeds at different rates for the 1G and the 2G populations. When we measure the degree of energy equipartition without separating the two populations, the combination of the dynamical properties of the two populations may lead to anomalous trends which are not representative of the cluster's evolutionary history and  its present dynamical state. 
We choose the tangential equipartition profile of the \SIMI\ model in Figure \ref{fig:meq_prof} to show an example of how the equipartition of the total system can follow a different pattern than that within each population separately.

While in the inner and intermediate regions of the cluster the value of $\mu_{\rm T}$ for both populations combined is intermediate between that of the 1G and that of the 2G, in the outermost regions it differs significantly from those of each population and is negative. 
A negative value of the equipartition mass describes a mass-velocity dispersion trend in which the velocity dispersion increases with the mass, an anomalous behavior opposite to that expected to emerge from the normal evolution towards energy equipartition, but has been found in the outer regions of some models of single-population star clusters (\citealt{2021PaVe}, \citeyear{2022PaVe}, \citealt{2022LiVe}, \citealt{2023ArVe}, \citealt{2024PaHe}).

In this case, however, this effect is due to a combination of the differences between tangential velocity dispersion of the two populations and the radial variation of the fraction of 2G stars.
To visualize how this occurs, we plot the tangential velocity dispersion as a function of mass in the 85$-$95\% radial shell of the \SIMI\ model for each population separately, as well as both combined in Figure \ref{fig:meq_unmixed}.
The fraction of 2G stars in the same radial shell is over-plotted to show its dependence on the stellar mass.
The velocity dispersion of both populations combined falls in-between the dispersion of the two populations at low stellar masses, but, since the fraction of 2G stars decreases for increasing stellar masses, the combined velocity dispersion approaches that of the 1G population for higher stellar masses.

This results in a positive correlation between the velocity dispersion and the stellar mass, a trend fit by a negative equipartition mass, for a combination of populations that, individually, have little to no trend of velocity dispersion with stellar mass.

Note that this effect is also prevalent in populations that have measurable levels of energy equipartition if the velocity dispersion is different between the two populations and the mass-dependent spatial mixing is not complete, but the effect is most noticeable in cases such as the example shown in \SIMI\ model in the bottom panel of Figure \ref{fig:meq_prof}.

\subsubsection{Time evolution of the degree of energy equipartition}
In order to illustrate the evolutionary history leading to the present-day radial variation of the degree of energy equipartition, we plot the time evolution of $\mu$ (Figure \ref{fig:meq_evo_tot}), $\mu_{\rm R}$ (top panel of Figure \ref{fig:meq_evo_Rphi}), and $\mu_{\rm T}$  (bottom panel of Figure \ref{fig:meq_evo_Rphi}) measured at a few selected projected distances from the cluster's center, as well as the differences between the values for the 2G and 1G populations for the \SIMII\ model. 

As expected, the inner regions are those characterized by a more rapid evolution towards energy equipartition, and the rate of the evolution towards energy equipartition decreases at larger distances from the cluster's center.

It is interesting to note that while during the early evolutionary phases differences between the values of $\mu$ of the 1G and the 2G populations are present both in the inner and intermediate regions, as the system evolves these differences persist mainly in the intermediate regions while both the inner and outermost regions of the cluster are characterized by small or negligible differences. In general, the 2G population is characterized by a stronger degree of energy equipartition than the 1G population.

The analysis of the degree of energy equipartition calculated separately for the radial and tangential velocity dispersion components reveal additional interesting aspects of the dynamics of multiple-population clusters.
In general the differences between the 1G and 2G values of $\mu_{\rm R}$ are smaller than those between the values of $\mu_{\rm T}$.
In particular, the radial equipartition mass in the cluster's outermost regions (top right panel of Figure \ref{fig:meq_evo_Rphi}) of the 2G and the 1G are characterized by similar values during the entire cluster evolution. 

As for the degree of energy equipartition in the tangential velocity dispersion,  we find that during the early evolutionary phases, in the intermediate and outer regions, the 2G population is characterized by a negative equipartition mass.  Despite this initial trend, during the subsequent evolution,  stars from the inner regions continue their outward diffusion and the 2G population evolves towards equipartition more rapidly than the 1G.

We visualize the time evolution of the differences in the equipartition between the two populations of each component for all of our models in Figure \ref{fig:meq_diff_evo}. 
Note that the largest differences at 12 Gyr are those between $\mu_{\rm T}$ in the intermediate regions; in the innermost regions of each model, the differences between the values of the equipartition mass of 1G and 2G stars are larger in the early evolutionary phases but are mostly erased by 12 Gyr. All models show stronger differences in the tangential component at later times, although in a few cases the differences in the tangential components are similar to those found in the radial component.

To further explore the differences between the radial and tangential equipartition for each population, we plot the radial profile of the difference in the inverse of the equipartition mass ($\mu_{\rm R}-\mu_{\rm T}=1/m_{\rm eq, R} - 1/m_{\rm eq, T}$), in Figure \ref{fig:ani_prof} in all of our models at 12 Gyr. 
This plot shows how, generally, the level of equipartition is "isotropic" in the inner regions, but can be characterized by significant differences between the radial and the tangential components at larger radii.
The \SIMIII\ model shows the most extreme differences between these two components.

To shed light on the evolution of these differences, we plot the time evolution of $\mu_{\rm R}$ and $\mu_{\rm T}$ in multiple radial shells in all of our models in Figure \ref{fig:meqani_evo_tot}.
These plots show that the inner regions are not in general characterized by strong differences between these two components; in the intermediate and outer regions, during the early evolutionary phases the difference between the radial and the tangential components is positive (corresponding to a stronger degree of energy equipartition in the tangential velocity dispersion) but this difference decreses over time and becomes negative for clusters reaching a more advanced dynamical age.
These differences are more significant in the 2G population than in the 1G population.

\section{Conclusions}
In this paper, we have studied the evolution towards energy equipartition in multiple-population clusters.
Our investigation is based on a set of Monte Carlo simulations exploring the evolution of clusters with different initial number of stars, structural and kinematic properties.
After a  brief overview of the general spatial and kinematic properties of the two populations and the level of dynamical mixing reached after 12 Gyr of evolution (see Figures \ref{fig:RadialMassmixing}-\ref{fig:TimeKinMix}), we focused our attention on the characterization of the energy equipartition for 1G and 2G stars and on the implications of the initial differences between the 1G and 2G properties for their evolution towards energy equipartition.
Our main conclusions are the following:

\begin{itemize}
\item Evolution towards energy equipartition is more rapid for the 2G population, and 2G stars are, in general, characterized by a stronger degree of energy equipartition than 1G stars at 12 Gyr
(it is interesting to note that a small difference between the degree of equipartition of 1G and 2G stars consistent with this trend was reported for $\omega$ Cen in \citealt{2018BeLi}).
Even systems where the two populations are essentially spatially mixed at 12 Gyr may still be characterized by some differences between the 2G and 1G degree of energy equipartition (see the \SIMIII\ model in Figures \ref{fig:MassMixSynth}, \ref{fig:DispMixSynth}, and \ref{fig:meq_prof}).

\item We have calculated the degree of energy equipartition using the total velocity dispersion or the radial and tangential components of the velocity dispersion separately. The evolution towards energy equipartition is "anisotropic" and proceeds at different rates in the tangential and radial directions. The anisotropy in energy equipartition is more prominent in the 2G population and in the intermediate and outer regions of the cluster.

\item Differences between the 1G and 2G equipartition are stronger at intermediate distances from the cluster's center and when calculated using tangential velocity dispersion (see Figure \ref{fig:meq_prof}, and Figures \ref{fig:meq_evo_Rphi}-\ref{fig:meqani_evo_tot}).

\item During the early evolutionary phases and in the clusters' outermost regions, the 2G population may develop negative values of the equipartition mass in the tangential component of the velocity dispersion; this corresponds to a trend between velocity dispersion and stellar mass that is in the opposite direction to that of clusters evolving towards energy equipartition. 
After these early phases, however, the 2G tangential component of the equipartition mass ultimately evolves more rapidly and further towards energy equipartition than the 1G population or the 2G radial component of the equipartition mass (see Figures \ref{fig:meq_evo_tot}, \ref{fig:meq_evo_Rphi}).
    
    \item Differences in the degree of energy equipartition of the 1G and the 2G populations concurrent with a dependence of the degree of mixing on the stellar mass may lead to an apparent anomalous dependence of the tangential velocity dispersion on the stellar mass in the total (1G+2G) population (see Figure \ref{fig:meq_unmixed}). 
\end{itemize}

In future investigations, we will further extend the investigation presented here to consider a broader range of initial conditions (e.g. exploring different initial 1G and 2G relative concentrations and kinematics, initial 1G-to-2G mass ratios, primordial binary fraction and black hole retention fractions) and carry out a comprehensive study of the dependence of the evolution towards energy equipartition of the 1G and 2G populations on the clusters’ initial dynamical properties.

\label{chap6:Conclusions}

\section*{Acknowledgements}
This research was supported in part by Lilly Endowment, Inc., through its support for the Indiana University Pervasive Technology Institute. AB and EV acknowledge support from STScI grants GO-15857 and AR-16157.
MG, AH, and AA acknowledge support by the Polish National Science Centre (NCN) grant 2021/41/B/ST9/01191.
AA acknowledges support for this research from project No. 2021/43/P/ST9/03167 co-funded by the Polish National Science Center (NCN) and the European Union Framework Programme for Research and Innovation Horizon 2020 under the Marie Skłodowska-Curie grant agreement No. 945339. For the purpose of Open Access, the authors have applied for a CC-BY public copyright license to any Author Accepted Manuscript (AAM) version arising from this submission.
TZ acknowledges funding from the European Union’s Horizon 2020 research and innovation programme under the Marie Skłodowska-Curie Grant Agreement No. 101034319 and from the European Union – NextGenerationEU.

\section*{DATA AVAILABILITY STATEMENT}
The data presented in this article may be shared on reasonable request to the corresponding author.

\clearpage

\clearpage
\bibliographystyle{mnras}
\bibliography{bib}
\end{document}